\documentclass[a4paper,11pt]{article}
\usepackage{enumitem}
\usepackage{setspace}
\usepackage{pos_modified}
\usepackage{fancyhdr}
\usepackage{subfig}
\usepackage{lineno}
\usepackage{cancel}
\usepackage{natbib}
\usepackage{multirow}
\usepackage{graphicx}

\newcommand{\Deg}{$^{\circ}$}
\renewcommand{\paragraph}[1]{\noindent \textbf{#1}}

\title{The UHECR-FR0 Radio Galaxy Connection: A Multi-Messenger Study of Energy Spectra/Composition Emission and Intergalactic Magnetic Field Propagation}

\ShortTitle{UHECR-FR0 Radio Galaxy Connection: A Multi-Messenger Study}

\author*[1]{J.P. Lundquist}
\emailAdd{jplundquist@gmail.com}
\author[2, 3]{L. Merten}
\author[1]{S. Vorobiov}
\author[4]{M. Boughelilba}
\author[4]{A. Reimer}
\author[4]{P.~Da~Vela}
\author[5]{F. Tavecchio}
\author[5]{G. Bonnoli}
\author[5]{C. Righi}

\affiliation[1]{Center for Astrophysics and Cosmology (CAC), University of Nova Gorica,  Vipavska 13, SI-5000 Nova Gorica, Slovenia}
\affiliation[2]{Theoretical Physics IV, Plasma Astroparticle Physics, Faculty for Physics and Astronomy, Ruhr University Bochum, 44780 Bochum, Germany}
\affiliation[3]{Ruhr Astroparticle and Plasma Physics Center (RAPP Center), Germany}
\affiliation[4]{Universit\"at Innsbruck, Institut f\"ur Astro- und Teilchenphysik, Technikerstraße 25/8, 6020 Innsbruck, Austria}
\affiliation[5]{Astronomical Observatory of Brera, Via Brera 28, 20121 Milano, Italy}

\abstract{
This study investigates low luminosity Fanaroff-Riley Type 0 (FR0) radio galaxies as a potentially significant source of ultra-high energy cosmic rays (UHECRs). Due to their much higher prevalence in the local universe compared to more powerful radio galaxies (about five times more than FR-1s), FR0s may provide a substantial fraction of the total UHECR energy density. To determine the nucleon composition and energy spectrum of UHECRs emitted by FR0 sources, simulation results from CRPropa3 are fit to Pierre Auger Observatory data. The resulting emission spectral indices, rigidity cutoffs, and nucleon fractions are compared to recent Auger results. The FR0 simulations include the approximately isotropic distribution of FR0 galaxies and various intergalactic magnetic field configurations (including random and structured fields) and predict the fluxes of secondary photons and neutrinos produced during UHECR propagation through cosmic photon backgrounds. This comprehensive simulation allows for investigating the properties of the FR0 sources using observational multi-messenger data.
}

\begin{document}
\thispagestyle{fancy}
\maketitle

\section{Introduction}
\label{sec:intro}
Ultra-high-energy cosmic rays (UHECRs) exceeding $10^{18}$ eV are the universe's most energetic particles, with astrophysical origins that remain uncertain. The Pierre Auger Observatory (Auger), a large area ($\sim$3000 km$^2$) hybrid detector located in Argentina, provides the most precise and high-statistics dataset of UHECR events. Auger analyses of the arrival direction distribution at large angular scales point to an extragalactic origin of UHECRs at energies E$\geq$8 EeV~\cite{Golup_ICRC2023}. A recent combined fit of the UHECR energy spectrum, shower maxima (X$_{\text{max}}$) distributions, and arrival directions supports an astrophysical model comprising homogeneous background sources and an adaptable addition from nearby source candidates~\cite{arXiv:2305.16693v1}. For instance, if these nearby sources are a selection of 44 starburst galaxies they could contribute up to a $\sim$20\% flux fraction at 40~EeV~\cite{arXiv:2305.16693v1}.



Low luminosity Fanaroff-Riley (FR0) radio galaxies have been identified as promising candidates for the primary sources of UHECRs \cite{Baldi:2017gao}. Our previous research has shown that FR0s can accelerate UHECRs to the maximum observed energies via a hybrid Fermi-I pre-acceleration with gradual shear acceleration \cite{Merten:2021brk}. Due to FR0s higher local prevalence compared to more energetic FR-1 or FR-2 radio galaxies (approximately five times more abundant than FR-1s at redshifts z$\leq$0.05), they could significantly contribute to the total UHECR energy density.

This study investigates FR0 radio galaxies as potential UHECR sources using multi-messenger observations. Simulations of FR0 isotropic UHECR emission and their intergalactic propagation to Earth through plausible magnetic fields using CRPropa3 \cite{AlvesBatista:2016vpy} are fit to Auger data~\cite{Yushkov:2020nhr, Deligny:2020gzq}. We estimate the FR0 emitted UHECR mass composition and energy spectra and the flux of cosmogenic secondary photons and neutrinos resulting from UHECR interactions with cosmic photon backgrounds.

\section{Methodology}
\label{sec:method}

\label{ssec:CRPropa3}
{\bf Intergalactic UHECR Propagation:} The CRPropa3 (v3.1.6) framework~\cite{AlvesBatista:2016vpy,Merten:2017mgk} is employed to simulate proton, helium, nitrogen, silicon, and iron UHECR primaries propagation through the intergalactic medium. 
Simulated UHECR interactions with the intergalactic photon backgrounds CMB, IRB (Gilmore12 model \cite{10.1111/j.1365-2966.2012.20841.x}), and URB (Protheroe96 model \cite{Protheroe:1996si}) account for photo-pion production, nuclear photodisintegration, Bethe-Heitler and gamma-gamma pair-production (single, double, and triplet), and inverse Compton scattering. Additionally, redshift adiabatic cooling and unstable nuclear decays are considered.

\begin{figure}[h!]
    \centering
    \vspace{-1.6em}
    \hspace{-0.4cm}
    \subfloat[Subfigure 1][]{
    \includegraphics[width=.35\linewidth]{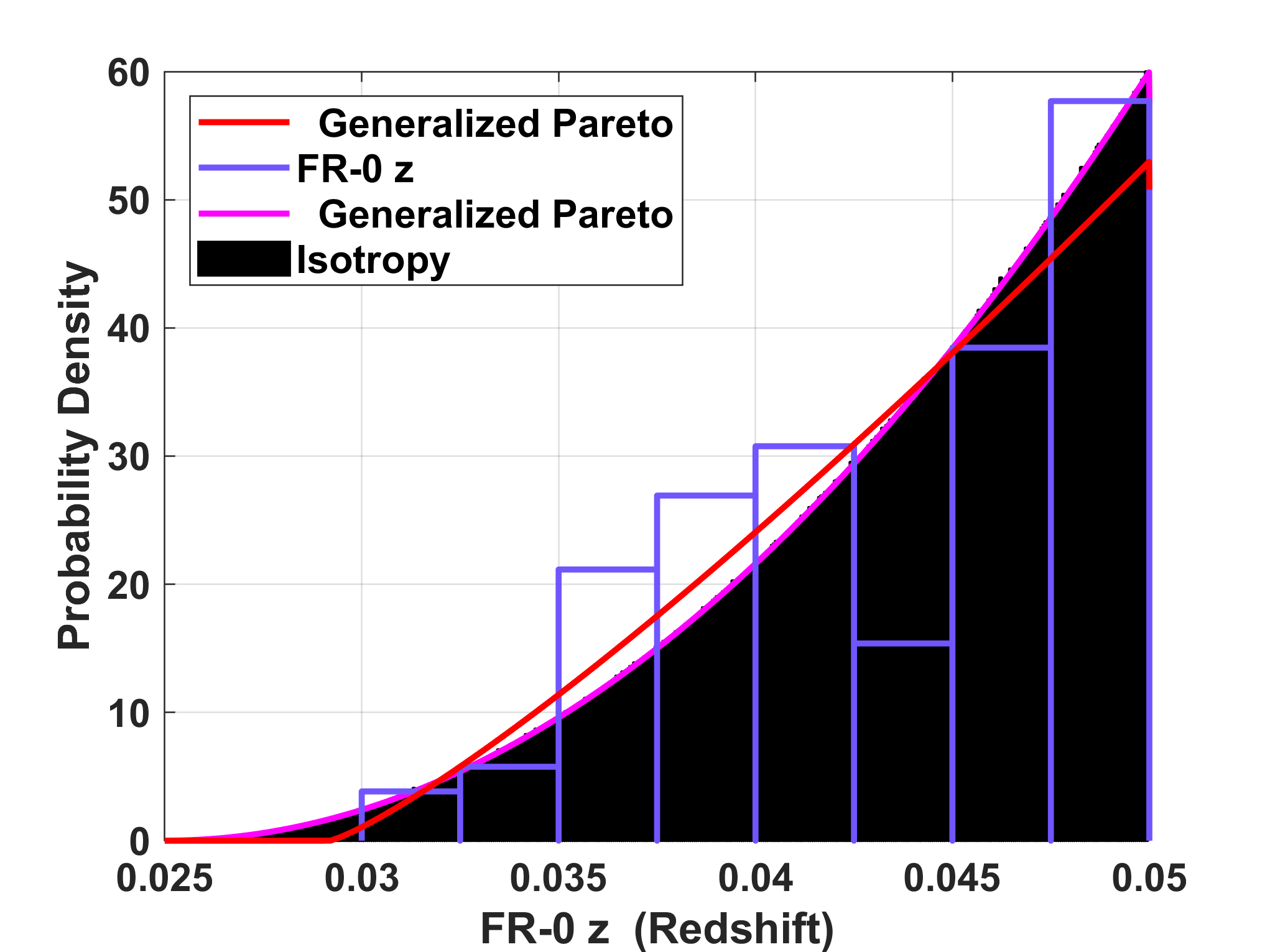}
    \label{fig:zhist}}%
    \hspace{-0.5cm}
    \subfloat[Subfigure 2][]{
    \includegraphics[width=.35\linewidth]{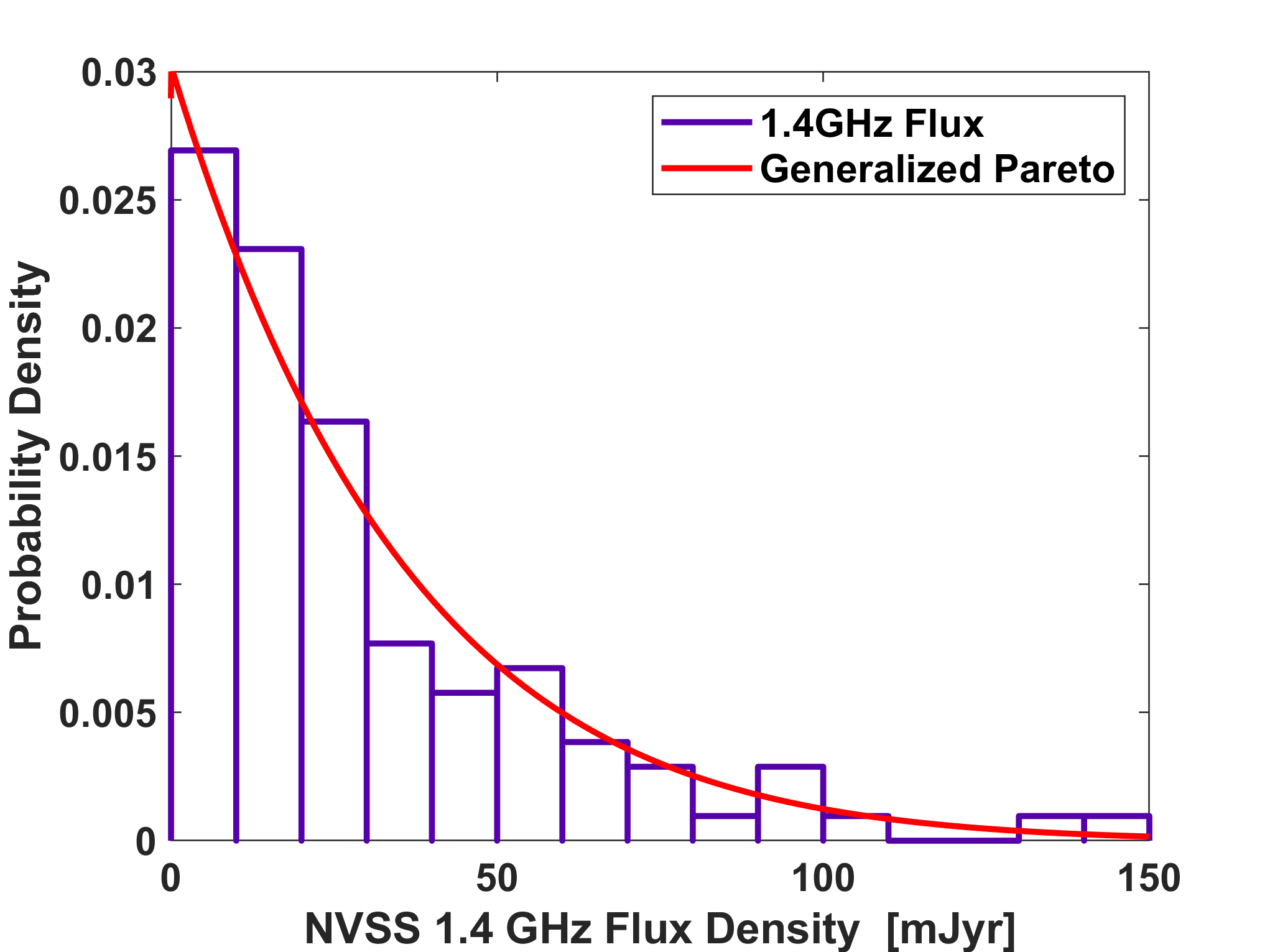}
    \label{fig:fluxhist}}%
    \hspace{-0.69cm}
    \subfloat[Subfigure 3][]{
    \includegraphics[width=.35\linewidth]{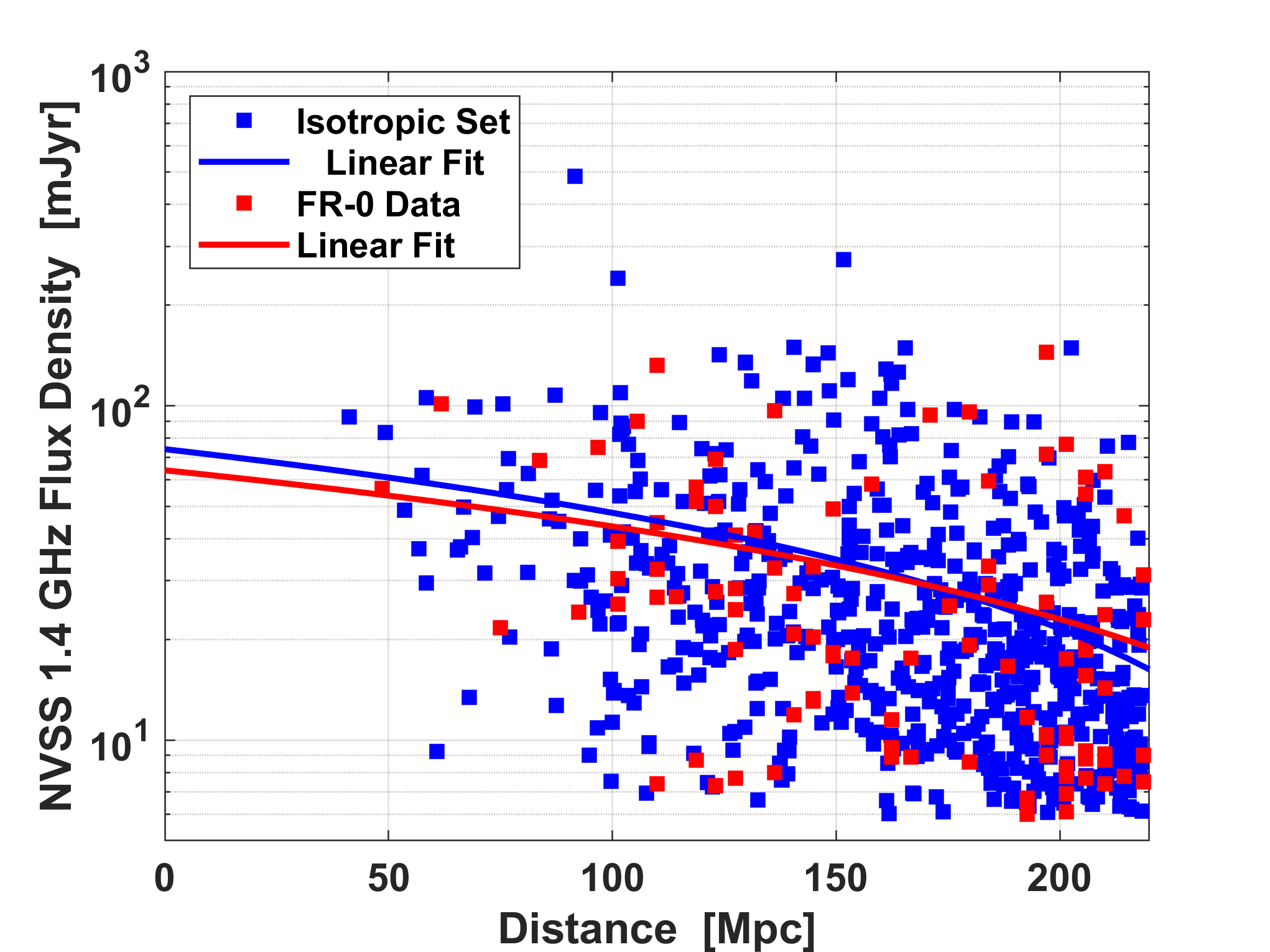}
    \label{fig:correlation}}
    \caption{(a) Simulated FR0 redshift, generated by fit to catalog data~\cite{Baldi:2017gao}. 
    (b) Relative simulated FR0 UHECR flux, proportional to the radio output distribution generated by fit to NVSS data~\cite{Baldi:2017gao}. (c) Local universe source evolution modeled by preserving redshift/flux correlation (Kendall corr. $\tau_b$: -0.28, p-val:~4.6$\times10^{-5}$)~\cite{Baldi:2017gao}.}
    \label{fig:distributions}
\end{figure}

\label{ssec:isotropic}
{\bf Isotropic FR0 Simulation:} The simulated FR0 are upsampled from the well-observed sky portion of the FR0CAT catalog~\cite{Baldi:2017gao}, ranging from -45\Deg to 45\Deg SGB and 60\Deg to 120\Deg SGL in supergalactic coordinates. 
Within this 11.79\% sky section, 76 out of 104 known FR0s are present, resulting in 645 upsampled FR0s with an isotropic pointing-direction distribution within a redshift $z<0.05$ distance (the maximum FR0CAT distance). These maintain the data redshift distribution (Figure~\ref{fig:zhist}) and have a relative flux proportional to the radio output distribution (Figure~\ref{fig:fluxhist}). Additionally, the ranked and linear correlation between radio (a source jet power measure)/UHECR flux and redshift (Figure~\ref{fig:correlation}) is preserved to model the local universe source evolution~\cite{Baldi:2017gao}. Each source emits flux isotropically, approximating that FR0s possess relatively slow bulk flows.


The CRPropa3 particle observer is a 200~kpc sphere and intersecting nuclei are recorded for energies E>10$^{18.6}$~eV. Particles travelling further than 4~Gpc or intersect a sphere 2.7~Gpc in radius are discarded. Observed secondary neutrinos with energies above 100~TeV are recorded. Secondary photons with energies above 100~MeV are immediately recorded if their path intersects the observer, and are then propagated using the \verb|DintElecaPropagation| module~\cite{Settimo:2013tua, Lee:1996fp}.

\begin{figure}[h!]
    \centering
    \vspace{-1em}
    \includegraphics[width=.45\textwidth]{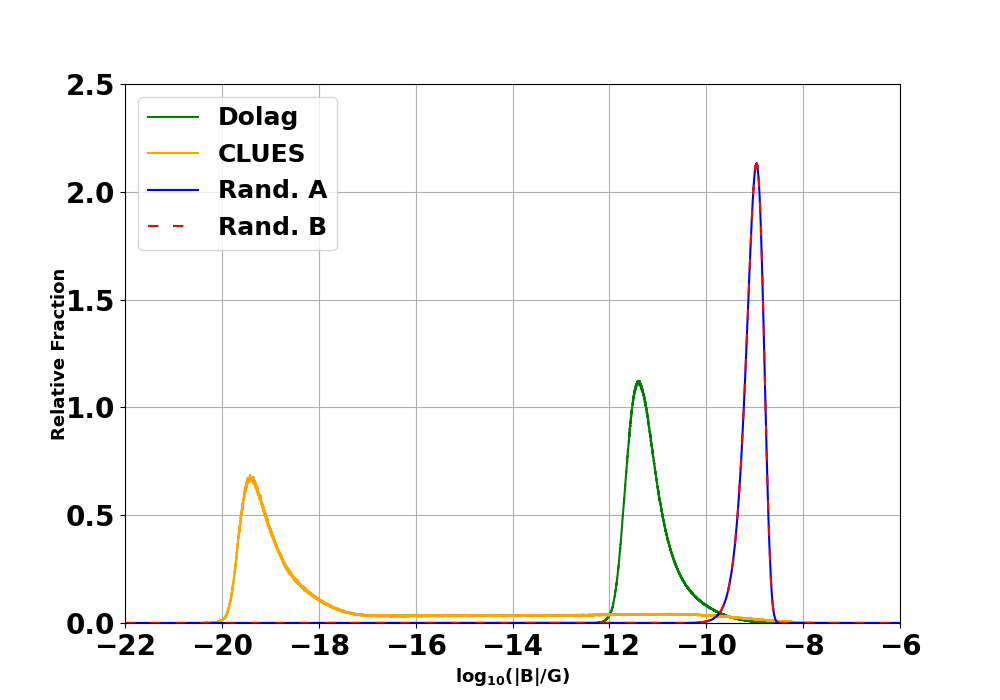}
    \caption{Magnetic field strength distributions for the $\langle B\rangle$=1~nG random fields (Rand A.: $\langle l_{\text{corr}} \rangle$=234~kpc and Rand B.: $\langle l_{\text{corr}} \rangle$=647~kpc), Dolag et al. structured field~\cite{Dolag_2005}, and CLUES 'astrophysical1R' field~\cite{Hackstein:2017pex}.}
    \label{fig:field}
\end{figure}

\label{ssec:fields}
{\bf Intergalactic Magnetic Fields:} We simulate magnetic field deflections using four magnetic field models and a no field scenario. Two fields are turbulent (via \verb|initTurbulence| module) with a mean strength of 1~nG and length scales from 60~kpc to 1~Mpc (or 3~Mpc) and a Kolmogorov power spectrum for an average correlation length of 234~kpc (or 647~kpc). The third model uses the Dolag et al.~\cite{Dolag_2005} structured field, while the fourth employs the Hackstein et al.\ (CLUES) 'astrophysical1R' structured field~\cite{Hackstein:2017pex}. Figure~\ref{fig:field} shows the magnetic field strength distributions.


\label{sec:fitting}
{\bf Composition and Energy Spectrum Combined Fit:} To find the simulated FR0 UHECR composition and energy spectrum emission that best matches Auger data~\cite{Yushkov:2020nhr, Deligny:2020gzq}, we minimize the sum $\chi^2$ per degree of freedom for composition and energy spectrum. 
Four fit parameters correspond to emitted proton, helium, nitrogen, silicon, and iron fractions over the whole energy range, constrained to a sum of one. The remaining parameters include the emitted spectral index $\gamma$, rigidity-dependent exponential cutoff~\cite{PierreAuger:2016use}, maximum particle trajectory cut-off, and energy spectrum normalization. The cost function is the sum $\chi^2$ per degree of freedom for composition and energy spectrum comparisons between simulation and data:
\begin{equation}
C =\sum{\chi_E^2/\text{dof}_E}+\sum{\chi_C^2/\text{dof}_C}
\label{eq:chi2}
\end{equation}

Equation~\ref{eq:chi2} is minimized using the SciPy library~\cite{2020SciPy-NMeth} with repeated stochastic Differential Evolution global minimization~\cite{Storn1997}, polished by deterministic local search. For redundancy the result is supplied to a stochastic Dual Annealing global minimization~\cite{XIANG1997216} and a final local search.

Uncertainties on the fit parameters are calculated as bootstrapped 68.27\% confidence intervals~(1$\sigma$ Gaussian equivalent) around the best fit values. The bootstrap samples include random sampling with replacement for the simulations and random Gaussian offsets to data using the total systematic and statistical uncertainties. In this way, uncertainties on the fitting method, simulation and data statistics, and systematic data uncertainties are taken into account.

\section{Results}
\label{sec:results}

\begin{figure}[h!]
    \centering
    \subfloat[Subfigure 1][]{
    \includegraphics[width=.45\textwidth]{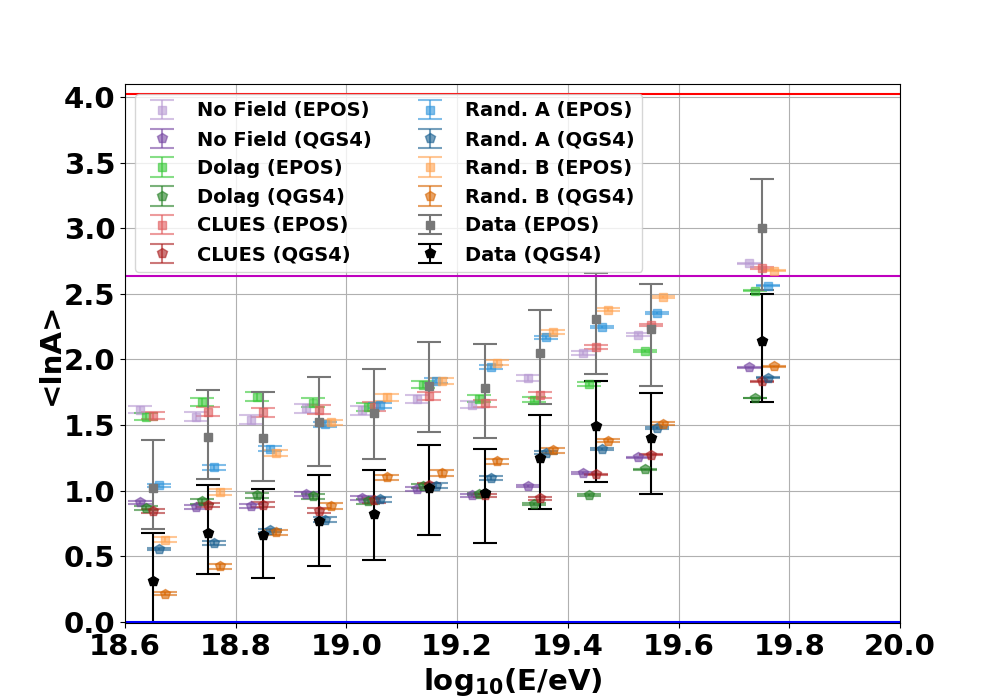}
    \label{fig:AmeanFit}}%
    \subfloat[Subfigure 2][]{
    \includegraphics[width=.45\textwidth]{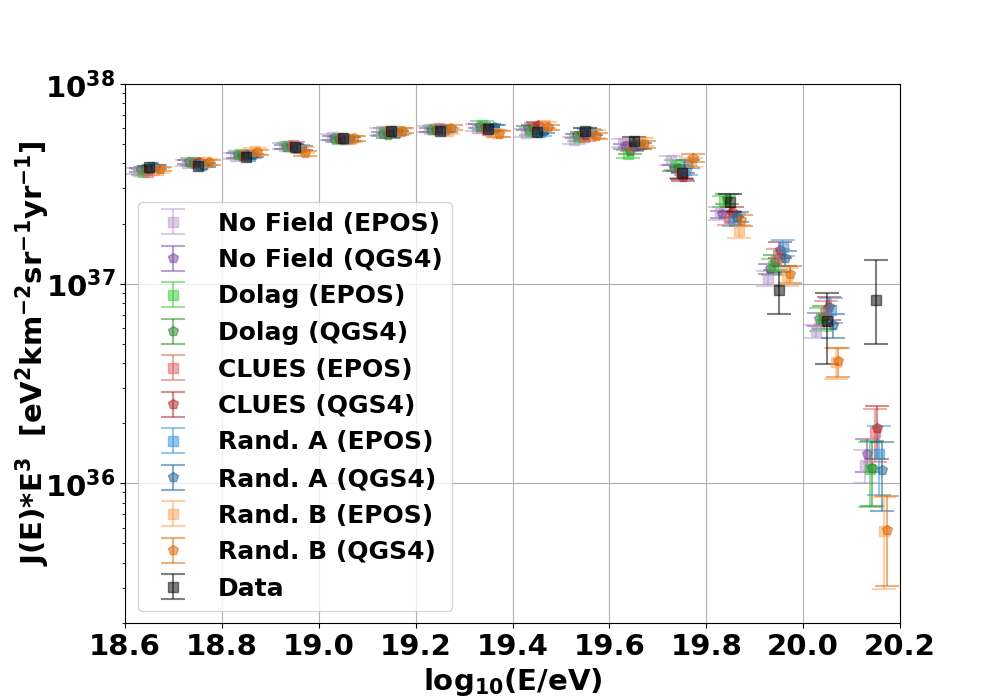}
    \label{fig:EspectFit}}%
    \caption{Combined fit results for all simulated intergalactic mediums and EAS models compared to Auger data~\cite{Yushkov:2020nhr, Deligny:2020gzq}. Slight offsets are applied in the x-axis for improved visibility. (a) Mean log mass number $\langle$lnA$\rangle$. Horizontal lines are blue:proton, magenta:nitrogen, and red:iron. (b) Energy spectra multiplied by E$^3$.}
    \label{fig:Fits}
\end{figure}

The combined fit results for all five simulated intergalactic mediums and two  extensive air-shower (EAS) models (EPOS-LHC and QGSJETII-04) compared to Auger data~\cite{Yushkov:2020nhr, Deligny:2020gzq} are shown in Figure~\ref{fig:Fits}. For the mean log mass number $\langle$lnA$\rangle$ (Figure~\ref{fig:AmeanFit}) it can be seen that the 1~nG random fields are more capable of fitting the lower energy, lighter composition bins. None of the models can account for the highest energy bins of the energy spectra~(Figure~\ref{fig:EspectFit}). FR0s are not expected to be a significant contributor at the most extreme energies.

The fit parameters are presented in Table~\ref{tab:params} along with corresponding~1$\sigma$ equivalent confidence intervals. For each EAS model and magnetic field this table shows the total $\Sigma \chi^2/\text{dof}$, emission spectral index~($\gamma$), exponential rigidity cutoff~(log$_{10}$($R_{\text{cut}}$)), trajectory cutoff~($D_{\text{cut}}$), and energy spectrum normalization~($n$).

\renewcommand{\arraystretch}{1.2}
\begin{table}[h]
\centering
\small
\rotatebox{0}{
\begin{tabular}{|c|c|c|c|c|c|c|}
\hline
\textbf{Field} & \textbf{Model} & \textbf{$\mathbf{\Sigma\chi^2/\text{dof}}$} &  \textbf{$\mathbf{\gamma}$} & \textbf{log$\mathbf{_{10}}$(R$\mathbf{_{cut}}$)} & \textbf{D$\mathbf{_{cut}}$} & \textbf{$\mathbf{n}$}\\
\hline
\multirow{2}{*}{No Field} & EPOS & 3.007 & 2.57$^{+0.03}_{-0.08}$ & 19.34$^{+0.15}_{-0.06}$ & 219$^{+0}_{-0}$ & 1.335$^{+0.009}_{-0.004}$\\
\cline{2-7}
 & QGS4 & 3.916 & 2.58$^{+0.02}_{-0.02}$ & 19.42$^{+0.03}_{-0.02}$ & 219$^{+0}_{-0}$  & 1.339$^{+0.005}_{-0.004}$\\
\hline
\multirow{2}{*}{Dolag} & EPOS & 4.430 & 2.63$^{+0.02}_{-0.11}$ & 19.43$^{+0.06}_{-0.12}$ & 226$^{+15}_{-0}$ & 1.335$^{+0.008}_{-0.007}$\\
\cline{2-7}
 & QGS4 & 4.304 & 2.60$^{+0.04}_{-0.04}$ & 19.48$^{+0.04}_{-0.03}$ & 230$^{+17}_{-4}$ & 1.341$^{+0.005}_{-0.004}$\\
 \hline
\multirow{2}{*}{CLUES} & EPOS & 3.178 & 2.63$^{+0.04}_{-0.07}$ & 19.37$^{+0.06}_{-0.07}$ & 219$^{+0}_{-0}$ & 1.333$^{+0.009}_{-0.004}$\\
\cline{2-7}
 & QGS4 & 3.774 & 2.65$^{+0.02}_{-0.05}$ & 19.45$^{+0.04}_{-0.02}$ & 219$^{+0}_{-1}$ & 1.341$^{+0.005}_{-0.006}$\\
 \hline
\multirow{2}{*}{Rand.A} & EPOS & 2.010 & 2.67$^{+0.22}_{-0.12}$ & 19.57$^{+0.18}_{-0.04}$ & 424.0$^{+111}_{-191}$ & 1.342$^{+0.005}_{-0.003}$\\
\cline{2-7}
 & QGS4 & 1.965 & 2.64$^{+0.10}_{-0.15}$ & 19.54$^{+0.03}_{-0.06}$ & 256$^{+161}_{-28}$ & 1.342$^{+0.005}_{-0.004}$\\
 \hline
\multirow{2}{*}{Rand.B} & EPOS & 4.314 & 2.92$^{+0.14}_{-0.30}$ & 19.62$^{+0.49}_{-0.05}$ & 401$^{+83}_{-116}$ & 1.345$^{+0.005}_{-0.005}$\\
\cline{2-7}
 & QGS4 & 4.472 & 2.80$^{+0.11}_{-0.11}$ & 19.58$^{+0.07}_{-0.04}$ & 281$^{+36}_{-49}$ & 1.344$^{+0.006}_{-0.003}$\\
\hline
\end{tabular}
}
\caption{The FR0 combined fit results total sum chi-square per degree of freedom, spectral index $\gamma$, exponential rigidity cutoff (log$_{10}$($R_{\text{cut}}$)), trajectory cutoff ($D_{\text{cut}}$), and spectrum normalization for all models.}
\label{tab:params}
\end{table}

The no field, CLUES, and Rand.A ($\langle B \rangle$=1 nG, $\langle l_{\text{corr}} \rangle$=234~kpc) magnetic field models generally exhibit the best fits. For the two EAS models, QGSJETII-04 has a higher $\chi^2$ value more frequently than EPOS-LHC. The best fit is the 1 nG random field (Rand.A) for either EAS model.

Overall, the table suggests that magnetic field strength impacts the cosmic-ray source emission energy spectrum. The emission spectral index tends to increase as magnetic field strength increases -- with generally softer emission spectra for QGSJETII-04 due to lighter required nuclei. The exponential rigidity cutoff (log$_{10}$($R_{\text{cut}}$)) also tends to increase with magnetic field strength -- with generally larger values for QGSJETII-04. The trajectory cutoff ($D_{\text{cut}})$) also tends to increase with magnetic field strength, which suggests that to match data, the maximum distance that cosmic rays travel through intergalactic space must increase as magnetic field strength increases. Finally, the energy spectrum normalization also tends to increase with magnetic field strength and is generally larger for QGSJETII-04 (due to the spallation of the heavier EPOS-LHC nucleons). This normalization result suggests that the number of cosmic rays emitted by sources at the highest energies must increase with magnetic field strength.


The FR0 nucleon emission fits for proton, helium, nitrogen, silicon, and iron are shown in Table~\ref{tab:nucleons} and Figure~\ref{fig:field}, along with their~1$\sigma$ equivalent confidence intervals. These results suggest that the magnetic field strength can have a significant impact on the composition of source emitted cosmic rays. As magnetic field strength increases, the proton percentage is relatively stable, helium and nitrogen nucleon percentages tend to increase, while the number of heavier nuclei tends to decrease. This heavy nucleon decrease is likely due to the fact that cosmic-ray deflection increases with magnetic field strength, and heavier nuclei are more easily deflected resulting in longer trajectories and a higher likelihood of observed secondaries from spallation.


\renewcommand{\arraystretch}{1.2}
\begin{table}[h]
\centering
\small
\rotatebox{0}{
\begin{tabular}{|c|c|c|c|c|c|c|}
\hline
\textbf{Field} & \textbf{Model} & \textbf{$\mathbf{f_H}$(\%)} & \textbf{$\mathbf{f_{He}}$(\%)} & \textbf{$\mathbf{f_N}$(\%)} & \textbf{$\mathbf{f_{Si}}$(\%)} & \textbf{$\mathbf{f_{Fe}}$(\%)} \\
\hline
\multirow{2}{*}{No Field} & EPOS & 90.0$^{+1.2}_{-4.6}$ & 0.0$^{+1.9}_{-0.0}$ & 0.0$^{+0.0}_{-0.0}$ & 8.7$^{+3.0}_{-1.6}$ & 1.3$^{+0.3}_{-0.5}$ \\
\cline{2-7}
 & QGS4 & 96.4$^{+0.7}_{-1.9}$ & 0.0$^{+0.9}_{-0.0}$ & 0.0$^{+0.0}_{-0.0}$ & 2.5$^{+0.8}_{-1.1}$ & 1.1$^{+0.3}_{-0.2}$ \\
\hline
\multirow{2}{*}{Dolag} & EPOS & 91.8$^{+1.6}_{-5.8}$ & 0.6$^{+5.1}_{-0.6}$ & 0.0$^{+0.0}_{-0.0}$ & 5.4$^{+2.4}_{-1.6}$ & 2.2$^{+0.6}_{-0.8}$ \\
\cline{2-7}
 & QGS4 & 93.3$^{+4.2}_{-4.9}$ & 4.2$^{+4.7}_{-4.2}$ & 0.0$^{+0.0}_{-0.0}$ & 1.0$^{+0.7}_{-0.8}$ & 1.5$^{+0.4}_{-0.3}$ \\
 \hline
\multirow{2}{*}{CLUES} & EPOS & 92.3$^{+1.0}_{-2.8}$ & 0.0$^{+0.7}_{-0.0}$ & 0.0$^{+0.0}_{-0.0}$ & 5.9$^{+1.6}_{-1.3}$ & 1.8$^{+0.4}_{-0.6}$ \\
\cline{2-7}
 & QGS4 & 97.4$^{+0.3}_{-4.0}$ & 0.0$^{+3.6}_{-0.0}$ & 0.0$^{+0.0}_{-0.0}$ & 1.2$^{+0.8}_{-0.7}$ & 1.4$^{+0.2}_{-0.3}$ \\
 \hline
\multirow{2}{*}{Rand.A} & EPOS & 71.7$^{+13.9}_{-21.0}$ & 19.5$^{+19.9}_{-0.0}$ & 6.6$^{+2.9}_{-2.6}$ & 1.2$^{+1.1}_{-0.6}$ & 1.0$^{+0.3}_{-0.6}$ \\
\cline{2-7}
 & QGS4 & 83.7$^{+9.4}_{-8.4}$ & 13.0$^{+8.9}_{-10.7}$ & 2.4$^{+1.3}_{-1.4}$ & 0.6$^{+0.3}_{-0.6}$ & 0.3$^{+0.2}_{-0.0}$ \\
 \hline
\multirow{2}{*}{Rand.B} & EPOS & 89.6$^{+1.3}_{-89.6}$ & 0.6$^{+86.4}_{-0.6}$ & 8.5$^{+1.7}_{-3.0}$ & 0.9$^{+2.9}_{-0.3}$ & 0.4$^{+0.2}_{-0.1}$ \\
\cline{2-7}
 & QGS4 & 95.3$^{+0.8}_{-0.8}$ & 0.0$^{+0.0}_{-0.0}$ & 4.2$^{+0.5}_{-1.0}$ & 0.3$^{+0.3}_{-0.2}$ & 0.1$^{+0.0}_{-0.1}$ \\
\hline
\end{tabular}
}
\caption{The FR0 combined fit results nucleon emission percentages for proton, helium, nitrogen, silicon, and iron primaries for all 10 models.}
\label{tab:nucleons}
\end{table}

\begin{figure}[t]
    \centering
    \includegraphics[width=0.9\textwidth]{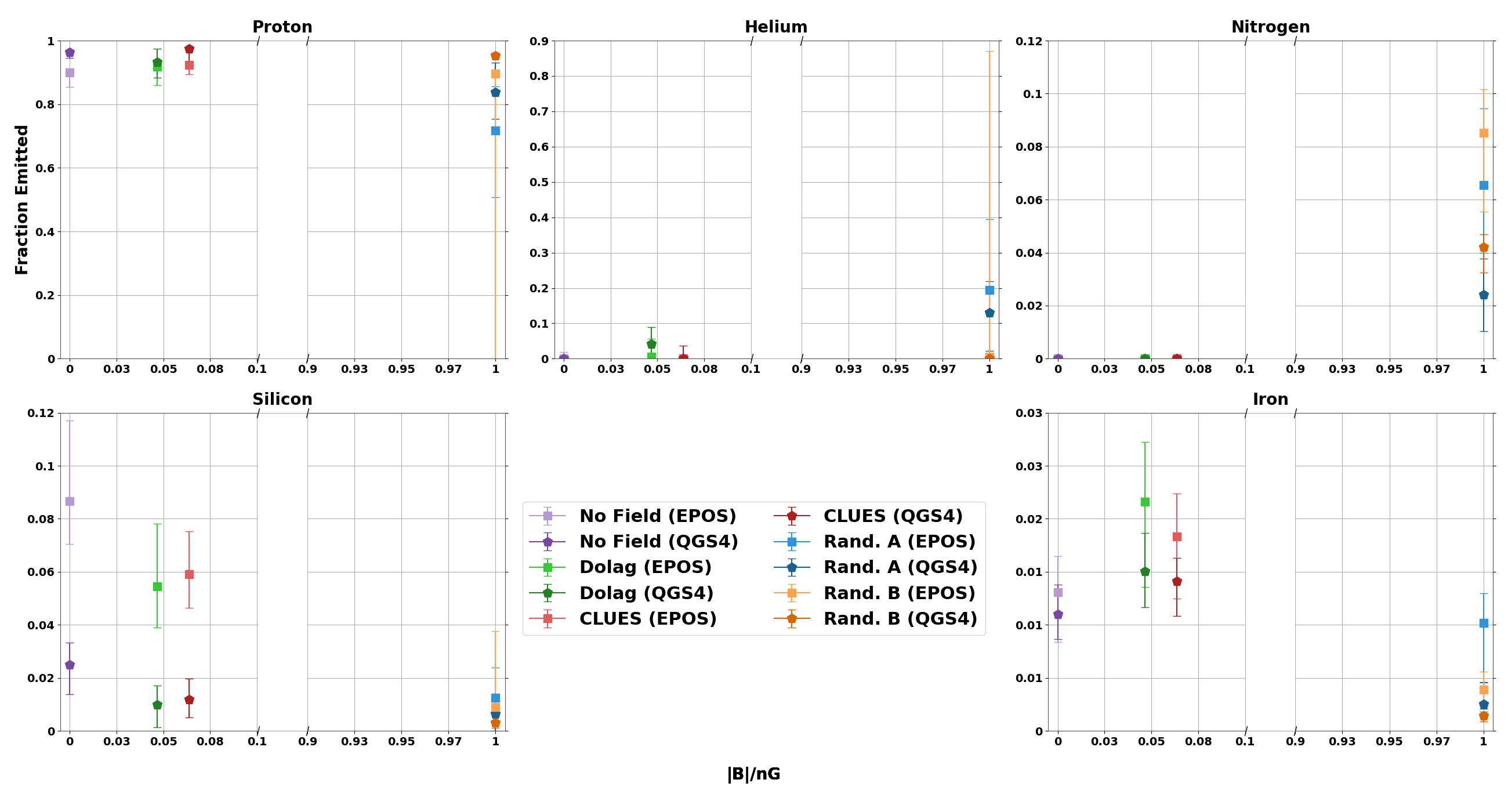}
    \caption{FR0 emitted nucleon fractions necessary to best fit the data for all 10 models versus magnetic field strength. Note the different y-axis limits and the broken x-axis.}
    \label{fig:nucleons}
\end{figure}

\begin{figure}[h!]
    \centering
    \subfloat[Subfigure 1][]{
    \includegraphics[width=.45\textwidth]{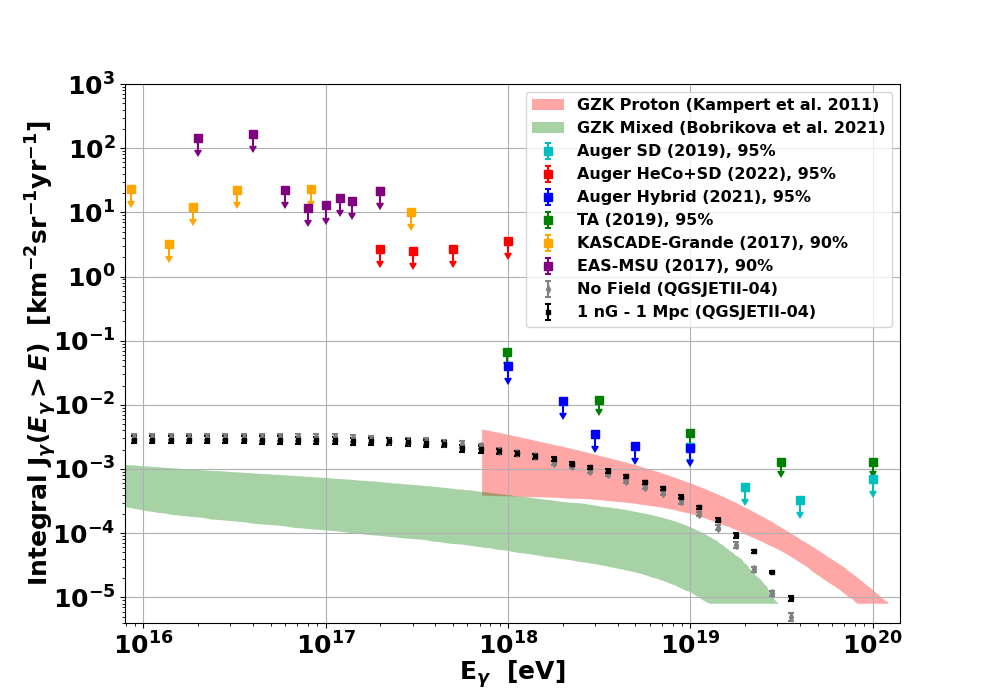}
    \label{fig:photon}}%
    \subfloat[Subfigure 2][]{
    \includegraphics[width=.45\textwidth]{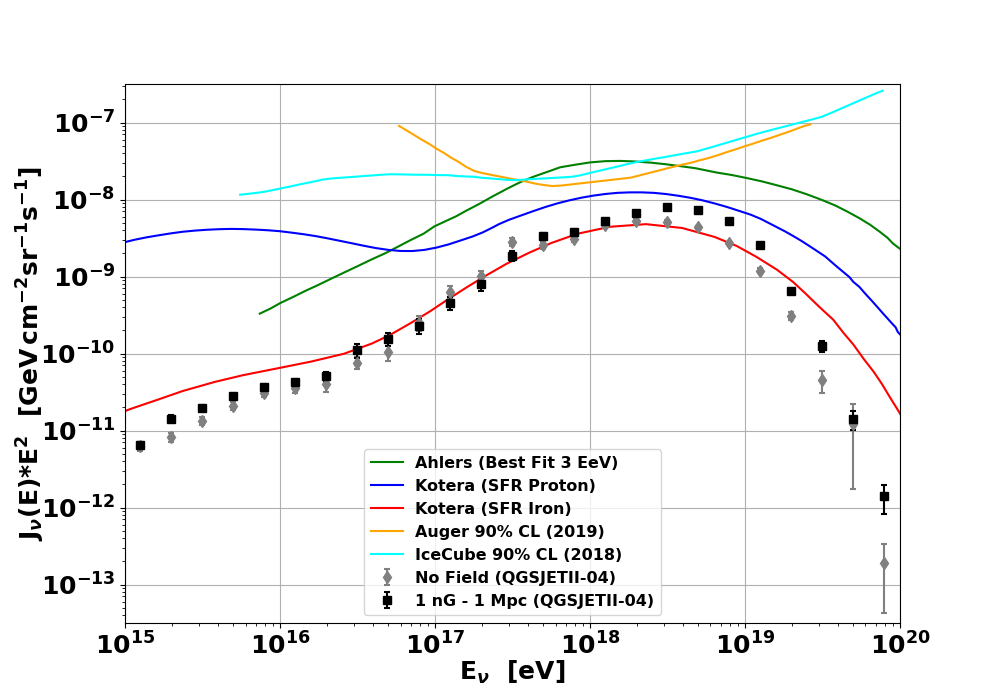}
    \label{fig:neutrino}}%
    \caption{Cosmogenic photon and neutrino spectra for the best fit $\langle B \rangle$=1 nG, $\langle l_{\text{corr}} \rangle$=234~kpc random magnetic field and QGSJETII-04 composition model (the no field case is grey). (a) Integral cosmogenic photon spectrum compared with two theoretical models and experimental upper limits as in~\cite{PierreAuger:2022uwd}. (b) All-flavor neutrino spectrum compared with three theoretical models and experimental upper limits~\cite{IceCube:2018fhm, PierreAuger:2019ens, 2010JCAP...10..013K}.}
    \label{fig:multimessenger}
\end{figure}

The cosmogenic integral photon and all-flavor neutrino spectra are shown in Figures~\ref{fig:photon} and~\ref{fig:neutrino} for the best fit model (compared to the no field scenario) -- $\langle B \rangle$=1 nG, $\langle l_{\text{corr}} \rangle$=234~kpc random field and QGSJETII-04 composition. This light nucleon emission model results in an integral photon flux mostly compatible with a pure proton GZK interaction prediction. Interestingly, the neutrino flux is similar to a pure iron prediction which may be partially due to the simulation constraint of relatively close FR0 sources (z<0.05). Given the wide range of theoretical predictions and current experimental upper limits the FR0 simulated flux is reasonable. Overall, in both spectra, it can be seen that a magnetic field results in a larger flux - particularly at the highest energies.

\section{Conclusion}


Fanaroff-Riley (FR0) radio galaxies have been shown to be capable of accelerating cosmic rays to ultra-high energies (E$>$$10^{18.6}$ eV)~\cite{Merten:2021brk, Baldi:2017gao}. In this study, we demonstrated that relatively close isotropic Fanaroff-Riley (FR0) radio galaxy emission (z<0.05) propagated through various extragalactic fields fit to published Pierre Auger Observatory UHECR composition and energy spectrum~\cite{Yushkov:2020nhr, Deligny:2020gzq} results in good agreements. Our analysis suggests a soft source energy spectrum with a spectral index of $\gamma$$\sim$-2.6 with a predominate proton emission. Furthermore, the estimated shapes of the secondary photon and neutrino fluxes created by UHECR interactions with cosmic photon backgrounds appear reasonable and well with in limits.

The findings from this study contribute to a better understanding of UHECR sources and their interactions with the intergalactic environment. By demonstrating a strong possible connection between FR0 radio galaxies and UHECRs, this research opens up new avenues for exploring the acceleration mechanisms and environmental factors influencing the production and propagation of UHECRs.

There are potential future directions to further investigate the UHECR-FR0 connection and improve simulation accuracy. These include: extrapolating FR0 sources to larger redshifts (z>0.05) and estimating the required FR0 cosmic-ray luminosity to provide further constraints on the UHECR emission properties of these sources. These and other FR0 avenues will help to further refine our understanding of UHECR acceleration, emission, and propagation.


\section*{Acknowledgements}
Financial support was received from the Austrian Science Fund (FWF) under grant agreement number I 4144-N27 and from the Slovenian Research Agency - ARRS (project no. N1-0111). LM acknowledges support from the DFG within the Collaborative Research Center SFB1491 "Cosmic Interacting Matters - From Source to Signal". MB has for this project received funding from the European Union’s Horizon 2020 research and innovation program under the Marie Sklodowska-Curie grant agreement No 847476. The views and opinions expressed herein do not necessarily reflect those of the European Commission. 
The Slovenian National Supercomputing Network provided extensive computing support.


\end{document}